\documentclass[11pt]{article}

\usepackage{pgf}

\usepackage[english]{babel}
\usepackage[utf8]{inputenc}
\usepackage{lmodern}
\usepackage[T1]{fontenc}
\usepackage{datetime}

\usepackage{amsmath,amssymb}
\usepackage{mathtools}
\usepackage{geometry}
\usepackage{array}
\usepackage{xspace}

\usepackage{amsmath,amsfonts,amssymb,amsopn,amsthm}
\theoremstyle{plain}
\usepackage{cases}
\usepackage[all]{xy}

\usepackage{url,amssymb}

\newtheorem{proposition}{Proposition}
\newtheorem{theorem}[proposition]{Theorem}
\newtheorem{lemma}[proposition]{Lemma}
\newtheorem{corollary}[proposition]{Corollary}

\theoremstyle{definition}
\newtheorem{definition}[proposition]{Definition}

\theoremstyle{example}

\begin{document}
\title{Vectorial Boolean functions and linear codes in the context of algebraic attacks }
\author{M. Boumezbeur{\thanks{USTHB, Laboratory of Algebra and number theory,  BP 32 El Alia, Bab Ezzouar, Algeria. E-mail: boumezbeur.m@live.fr}}, S. Mesnager{\thanks{University of Paris VIII (department of mathematics), University of Paris XIII (LAGA) and Telecom ParisTech, France. Email: smesnager@univ-paris8.fr}}, K. Guenda{\thanks{USTHB, Laboratory of Algebra and number theory,  BP 32 El Alia, Bab Ezzouar, Algeria. Email: ken.guenda@gmail.com}}}
\date{}
\maketitle

\begin{abstract}
In this paper we study the relationship between vectorial (Boolean) functions and cyclic codes in the context of algebraic attacks.
We first derive a direct link between the annihilators of a vectorial function (in univariate form) and certain $2^{n}$-ary cyclic codes (which we prove that they are LCD codes) extending results due to R\o njom and Helleseth. The knowledge of the minimum distance of those codes gives rise to a lower bound on the algebraic immunity of the associated vectorial function. Furthermore, we solve an open question raised by Mesnager and Cohen.  We also present some properties of those cyclic codes (whose generator polynomials determined by vectorial functions) as well as their weight enumerator. In addition we generalize the so-called algebraic complement  and study its properties. 

\end{abstract}

\section{Introduction}

Error correcting codes are widely studied by researchers and employed by engineers.  They have long been known to have applications in computer and communication systems, data storage
devices (starting from the use of Reed Solomon codes in CDs) and consumer electronics. On the other hand, substitution boxes (S-boxes) are fundamental parts of block ciphers. They play an important role in their robustness, by providing confusion.
Mathematically, S-boxes are vectorial (multi-output) Boolean
functions, that is, functions from the vector space $\mathbb{F}^{n}_{2}$ (of all binary
vectors of length $n$) to the vector space $\mathbb{F}^{m}_{2}$, for given positive integers
$n$ and $m$. These functions are called $(n,m)$-functions and
include the (single-output) Boolean functions (which correspond to the case $m
=1$). When they are used as S-boxes in block ciphers, their number $m$ of output bits equals or approximately equals the number $n$ of input bits. They can also be used in stream ciphers, with $m$ significantly smaller than $n$, in the place of Boolean functions to speed up the ciphers.  We recall that to resist against known attacks, the cryptographic Boolean and vectorial function  involving in the cryptosystems should be designed properly.  In particular Boolean functions should be balanced to resist against statistical attacks, high algebraic degree to resist against Berlekamp-Massey's attack, high order correlation immunity against correlation attacks (in the combiner model of stream ciphers) and should have high algebraic immunity to resist against algebraic attacks. For multi-output Boolean functions $F$, various measures of nonlinearity have been widely discussed in the literature from coding point of view (see for instance \cite{Liu-Mesnager}. In particular, it has been shown that their corresponding linear code  and their corresponding punctured code are related to some special linear codes. The aim of this paper is to exhibit further links between vectorial Boolean functions and linear codes. We are dealing with vectorial functions in the context of algebraic attacks \cite{Bat, CourMei, CourPie, MeiPasCar} (which principe goes back to Shanon's work \cite{SHAN}) which are among the most efficient attacks for secret key cryptosystems  introduced in 2003 by Courtois and Meier in \cite{CourMei} and Courtois and Pieprzyk in \cite{CourPie}. The concept of algebraic immunity of Boolean functions was proposed by Meier et al. in \cite{MeiPasCar}. Next, it has been generalized to vectorial functions by Armknecht and Krause in \cite{ArmKra}. The authors of \cite{CourM} proved that $\lceil\dfrac{n}{2}\rceil$ is the maximal value of the algebraic immunity of $n-$variable Boolean functions. In \cite{ArmKra} the authors gave an upper bound on the algebraic immunity of vectorial function with $n$ inputs and $m$ outputs. Some constructions of Boolean and vectorial functions with optimal algebraic immunity were proposed in \cite{ArmKra}, \cite{CaDaGuMa}, \cite{CarlFre} and \cite{ZhSoDuWe}. In \cite{CarlFre} Carlet and Feng gave a construction of an infinite class of Boolean functions with optimal algebraic immunity, their construction was obtained from the BCH bound of cyclic codes, this gives a link between the algebraic immunity of Boolean functions and cyclic codes.
Furthermore, by considering the univariate representation of Boolean functions, R\o njom and Helleseth \cite{HellRon} gave a direct link between the annihilators of Boolean functions and some $q-$ary cyclic codes. They proved that the algebraic immunity of Boolean functions is equal to the minimum distance of those codes. Another direct link between Boolean functions and cyclic codes was presented in \cite{Mesn}, where Mesnager and Cohen gave a lower bound on algebraic immunity of Boolean functions by studying the properties of some cyclic codes. In this paper we firstly derive a direct link between the annihilators of vectorial functions and some cyclic codes. The knowledge of the minimum distance of those codes gives rise to a lower bound on the algebraic immunity of the associated vectorial function. We present some properties of those cyclic codes for which their generator polynomials are determined by vectorial functions.

This paper is organized as follows. In section 2 we give some preliminaries. In section 3 we present a direct link between the annihilators of a vectorial functions and certain cyclic codes generalizing the link between Boolean functions and cyclic codes given in \cite{HellRon}. More specifically, we show that any annihilator of an $(n,m)$-function $F$ belong to an $2^n$-ary cyclic code with generator polynomial $G_F$ (Theorem 8). Next, we describe the set of zeros of such cyclic code $\mathcal C(F^{-1}(b))_{b\in\mathbb F_{2^m}}$ from which we deduce that the algebraic immunity is equal to the minimal weight-heigh of a codeword of $\mathcal C(F^{-1}(b))_{b\in\mathbb F_{2^m}}$ (Corollary 10). Furthermore, we extend in Subsection 3.2 the results given in \cite{Mesn} to $(n,m)$-functions $F$ and derive a lower bound on the lowest algebraic degree of nonzero annihilators of $F$ (Theorem 12). In subsection 3.3 we give a relation between the spectral immunity,  algebraic immunity of an $(n, m)$-vectorial function $F$ and the minimal weight of $\mathcal C(F^{-1}(b))_{b\in\mathbb F_{2^m}}$ (Theorem 13).
In section 4 we extend the notion of algebraic complement for vectorial functions and present several results. 
In particular the relationship between the annihilators of the $(n, m)$-function F and the annihilators of its algebraic complement is given. In Subsection 4.3, we study  the lowest algebraic degree of nonzero annihilators of Boolean and vectorial functions and answer to an open problem raised in \cite{Mesn}.
 In section 5  we show that the cyclic code $\mathcal C(F^{-1}(b))_{b\in\mathbb F_{2^m}}$ associate to the $(n,m)$-function $F$ is a linear code with complementary duals (abbreviated LCD) which is a linear code whose intersection with its dual is trivial \cite{mas92}. Further, we present  the parameters of $\mathcal C(F^{-1}(b))_{b\in\mathbb F_{2^m}}$ as well as some of its  properties. 
 \section{Preliminaries and notation}
Throughout this paper, $\vert E\vert$ will denote the cardinality of a finite set $E$.
\subsection{Some background on vectorial Boolean functions}

Given two positive integers $n$ and $m$, a mapping from $\mathbb{F}^{n}_{2}$ to $\mathbb{F}^{m}_{2}$, is called an $(n,m)$-vectorial function. Boolean functions correspond to the case where $m=1$. 
An $(n,m)$-function $F$ being given, the {\em coordinate
functions} of $F$ are the Boolean functions $f_1,\cdots,f_m$ defined by $F(x)=(f_1(x), \cdots,f_m(x))$ at every $x\in \mathbb{F}_2^n$. The {\em component functions} of $F$ are the linear combinations of its
coordinate functions, with non all-zero coefficients.\\
In cryptography, the most used representation of an $(n,m)$-function is the so-called Algebraic Normal Form (ANF):
\begin{displaymath} F(x_1,\cdots,x_n)=\sum_{u\in \mathbb{F}_2^n} c(u)
\left(\prod_{i=1}^{n} x_i^{u_i}\right),\quad c(u)\in \mathbb{F}_2^m.
\end{displaymath}
The {\em algebraic degree} $deg(F)$ of any $(n,m)$-function $F$ is
by definition the global degree of its ANF. It also equals the maximum algebraic degree
of the coordinate functions of $F$ or of its component functions.
Let $\mathbb{F}_{2^{n}}$ be the Galois field of order $2^n$. By identifying $\mathbb{F}^{n}_{2}$ to $\mathbb{F}_{2^{n}}$,
$(n,m)$-functions can be viewed  as functions from $\mathbb{F}_{2^{n}}$ to
$\mathbb{F}_{2^{m}}$. In this case, the component functions are the functions $Tr^{m}_{1}(v F(x))$ where $Tr^{m}_{1}$ is the {\em absolute trace} from $\mathbb{F}_{2^{m}}$ to $\mathbb{F}_{2}$: $Tr^{m}_{1}(x) = \sum_{i=0}^{m-1} x^{2^i}$, the function $(x,y) \mapsto Tr^{m}_{1}(x y)$ being an inner product in
$\mathbb{F}_{2^{m}}$.\\

Every mapping from $\mathbb{F}_{2^{n}}$ into $\mathbb{F}_{2^{n}}$ admits a unique univariate polynomial representation:
$$F(x) = \underset{i=0}{\overset{2^{n}-1}{\sum}} \delta_{i} x^{i}, \delta_{i} \in \mathbb{F}_{2^{n}}.$$

The algebraic degree of $F$ is equal to $\max_{j\, |\, c_j\neq
0}w_2(j)$ where we recall that $w_2(j)$ is the {\em 2-weight} of $j$, that is, the
number of nonzero coefficients $j_s$ in the binary expansion
$\sum_{s=o}^{n-1}j_s2^s$ of $j$.

For every integer $m$ dividing $n$, an $(n,m)$-function $F$ can be
viewed as a function from $\Bbb{F}_{2^n}$ to itself and, therefore
admits a unique univariate polynomial representation, which can be
represented in the form $Tr_{m}^n (\sum_{j=0}^{2^n-1}c_jx^j)$,
where $Tr_{m}^n(x)=x+x^{2^m}+x^{2^{2m}}+x^{2^{3m}}+\cdots
+x^{2^{n-m}}$ is the trace function from $\Bbb{F}_{2^n}$ to
$\Bbb{F}_{2^m}$ (but without uniqueness if we do not add
restrictions on the polynomial inside the brackets).

Given a Boolean function $f:\, \mathbb{F}_{2^{n}} \rightarrow \mathbb{F}_{2}$, its  support denoted by $supp(f)$ is defined as $supp(f):=\lbrace x \in \mathbb{F}_{2^{n}},\, f(x)=1 \rbrace$. 

\begin{definition}[Walsh Hadamard transform]

The Walsh Hadamard transform of $f:\, \mathbb{F}_{2^{n}} \rightarrow \mathbb{F}_{2}$ at $u \in \mathbb{F}_{2^{n}}$  is defined as   
\begin{equation}
W_{f}(u)= \underset{x \in \mathbb{F}_{2^{n}}}{\sum} (-1)^{f(x) +Tr_{1}^{n}(ux)}. \label{eq00}
\end{equation}
\end{definition}

\begin{definition}[Annihilator \cite{Feng}]
Let $H$ be a subset of $\mathbb{F}_{2}^{n}$, any function $g:\, \mathbb{F}_{2}^{n} \rightarrow \mathbb{F}_{2}$ of the set $I(H)=\lbrace g\neq 0, \, g(x)=0, \forall x \in H \rbrace$, is called an annihilator of $H$.
\end{definition}

Given an $(n,m)-$vectorial function $F$ in univariate form, with $m$ divides $n$, we shall use the following definition of annihilators of $F$:

\begin{definition}\cite{Nico}.
The Boolean function $g\neq 0$ is an annihilator of $F$ if $g(x)F(x)=0$, for all $x\in \mathbb{F}_{2^{n}}$, where for $b\in \mathbb{F}_{2^{m}}$, $g(x)=0$ for all $x\in F^{-1}(b):=\lbrace a\in \mathbb{F}_{2^n},\, F(a)=b \rbrace$.
\end{definition}

In \cite{ArmKra} Armknecht and Krause have introduced the algebraic immunity of $(n,m)$-functions and gave the following definition.

\begin{definition}[Algebraic immunity \cite{ArmKra}]
Let $F$ be an $(n,m)$-function. The algebraic immunity of $F$ is defined as follows
$$AI(F)=min \lbrace deg(g): g \neq 0; \, \exists b \in \mathbb{F}^{m}_{2}\mid g(x)=0,  \forall x \in F^{-1}(b)\rbrace$$
\end{definition}

We denote by $LDA(F)$ the lowest algebraic degree of nonzero annihilators of $F$.
For more about Boolean and vectorial functions, we send the reader to the excellent chapters \cite{CarlB,Carl}.

\subsection{Linear binary sequences and spectral immunity}
A binary sequence $s=(s_{t})_{t\geq0}$ is called linear recurring sequence of order $n$ over $\mathbb{F}_{2}$, if the sequence satisfies

$$s_{t+n}+ h_{n-1}\cdot s_{t+n-1} + h_{n-2}\cdot s_{t+n-2} + \cdots + h_{0}\cdot s_{t} = 0,$$

where $n$ is a positive integer, $t\geq0$ and $h_{0},\cdots,h_{n-1} \in \mathbb{F}_{2}$. The initial state vector is the state vector $s_{0}= (s_{0}, s_{1}, \cdots,s_{n})$.\\
The sequence $s$ is generated with the characteristic polynomial

$$h(x)=x^{n}+h_{n-1}x^{n-1}+\cdots+h_{1}x+h_{0}.$$

If $h(x)$ is with the least degree, then it is called the minimal polynomial of $s$ and its degree is the linear complexity of the sequence and is denoted by $lc(s)$.\\
If there exists a positive integer $T$ such that $s_{t+T}=s_{t}$, then the sequence is called periodic of period $T$.

\begin{definition}(Annihilator of a binary sequence \cite{WaChZh})
For a binary sequence $s$ of period $T$, the binary sequence $a \neq 0$ is also of period $T$ satisfying $a\cdot s=0$ is called an annihilator of $s$.
\end{definition}
\begin{definition}\label{def1}(Spectral immunity \cite{GoHellRon,HellRon})
The spectral immunity of the binary sequence $s$, denoted by $SI(s)$, is the lowest linear complexity of all nonzero binary annihilators of $s$, that is
$$SI(s)= \underset{\lbrace a \neq 0; \, a\cdot s=0 \rbrace}{min}lc(a).$$
\end{definition}
A filter generator is a key-stream generator composed of one LFSR. If $z_{t}$ is the sequence generated the filter generator, then the output sequence is given by 
\begin{equation*}
    s_{t}=f(z_{t+\gamma_{1}},z_{t+\gamma_{2}},\cdots,z_{t+\gamma_{n}}).
\end{equation*}

\subsection{Cyclic codes}
In following we recall some basic notations and definitions on cyclic codes. 
\begin{definition}(Cyclic codes)
A linear $[n,k]-$code $\mathcal{C}$ over $\mathbb{F}_{2}$ is called cyclic if for any codeword $c= (c_{0},\cdots,c_{n-1}) \in \mathcal{C}$, the vector $c= (c_{n-1},c_{0},\cdots,c_{n-2})$ is also a codeword of $\mathcal{C}$.\\
Let $A_{i}$ denote the number of codewords in $\mathcal{C}$ with Hamming weight $i$. The weight distribution of $\mathcal{C}$ is the sequence the $(A_{1}, A_{2}, \cdots, A_{n})$ and the weight enumerator is defined by $Q(z)=1+A_{1}z+A_{2}z^{2}+ \cdots +A_{n}z^{n}$.  
\end{definition}
The dual code $\mathcal{C}^{\perp}$ of a cyclic code $\mathcal{C}$ is the cyclic code defined by
\begin{center}
$\mathcal{C}^{\perp} =\lbrace v \in \mathbb{F}_{2}^{n}; \, v\cdot c=0,\, \forall c \in \mathcal{C} \rbrace$;
\end{center}
where $v\cdot c$ is the usual inner product of $v=(v_{1}, \cdots ,v_{n})$ and $c=(c_{1}, \cdots,c_{n})$ in $\mathbb{F}_{2}^{n}$,
\begin{center}
$v\cdot c= \underset{i=1}{\overset{n}{\sum}} v_{i}\cdot c_{i}$.
\end{center}

Let $S$ be a subset of $\mathbb{F}_{2^{n}}$, let $\mathcal{C}(S)$ be the set of all tuples $(a_{1}, \cdots, a_{2^{n}-1})$ of $\mathbb{F}^{2^{n}-1}_{2^{n}}$ such that $\underset{i=1}{\overset{2^{n}-1}{\sum}}a_{i}x^{i}=0$ for all $x \in S$. Then $\mathcal{C}(S)$ is a cyclic code of length $2^{n}-1$.\\

The generator polynomial $g(x)$ of a cyclic code $\mathcal{C}$ is a factor of $x^{n}-1$ in $\mathbb{F}_{2^{n}}[x]$ and every codeword in the code can be expressed as a multiple of $g$.\\ 
Let $p(x)=(x^{n}-1)/g(x)$, then the generator polynomial of $\mathcal{C}^{\perp}$ is
\begin{equation}
    g^{\perp} = x^{k}p(x^{-1})/p(0). \label{eq}
\end{equation} 
The defining set of the cyclic code $\mathcal{C}$ is the set $\lbrace i\in\mathbb{Z},\, g(\alpha^{i})=0 \rbrace$.
For more about cyclic codes, we send the reader to the excellent book \cite{MacSlo}.

\section{Vectorial functions and cyclic codes in the context of algebraic attacks}
In this section we  shall extend the results of \cite{HellRon} to  $(n,m)$-functions and we give bounds on the algebraic immunity and the spectral immunity of those functions.\\

\subsection{Algebraic immunity of vectorial functions and cyclic codes}
In \cite{HellRon} R\o njom and Helleseth gave a connection between Boolean functions and certain $2^{n}-$ary cyclic codes and then related the problem of the estimation of the algebraic immunity to the low-weight height codewords in cyclic codes. We recall that the weight height of a polynomial $g$ is defined to be
\begin{center}
$wh(g)=max \lbrace wt(i); \, c_{i} \neq 0 \rbrace$,
\end{center}
where $wt(i)$ is the Hamming weight of the vector $i$, defined as the number of its nonzero entries (see \cite{HellRon}). \\

The next result shows that any annihilator of a Boolean function $f$ belongs to certain $2^{n}-$ary cyclic code of length $2^{n}-1$ and gives an estimation of $Al(f)$.

\begin{theorem} \cite{HellRon} \label{proro}
Let $f$ be a Boolean function in univariate form, any annihilator of $f$ belongs to the $2^{n}$-ary cyclic code with generator polynomial $G_{f}$ such that
\begin{center}
$G_{f}= gcd (f(x)+1, x^{2^{n}-1}+1)$.
\end{center}

The algebraic immunity of $f$ is equal to the minimal weight-height of a codeword $g=\underset{i=0}{\overset{2^{n}-2}{\sum}} c_{i}x^{i}$ in the cyclic codes generated by $G_{f}$ and $G_{1+f}$.\\ 
\end{theorem}

In following we extend Theorem \ref{proro} to $(n,m)$-functions and derive a direct link between $(n,m)$-functions and $2^{n}-$ary cyclic codes.

\begin{theorem}\label{theo1}
Let $F$ be an $(n,m)$-function given in its univariate representation. Then any annihilator of $F$ belongs to the $2^{n}$-ary cyclic code with generator polynomial $G_{F}$ given by
\begin{equation}
    G_{F} = \underset{a \in \mathbb{F}_{2^{n}}^{*}}{\prod}gcd (F(x)-a, x^{2^{n}-1}+1). \label{vect}
\end{equation}
\end{theorem}

\begin{proof}
Let $g$ be an annihilator of $F$ such that $g(x)F(x)=0$ for all $x \in \mathbb{F}_{2^{n}}$. This holds for any $x \in \mathbb{F}^{*}_{2^{n}}$ and we obtain
\begin{center}
$g(x)F(x)\equiv 0 \, mod$ $(x^{2^{n}-1}+1),$
\end{center}
which implies
\begin{center}
$g(x)\equiv 0 \, mod$ ($\dfrac{x^{2^{n}-1}+1}{gcd(F(x), x^{2^{n}-1}+1})$.
\end{center}
Since the $(n,m)$-function $F(x)$ takes values in $\mathbb{F}_{2^{n}}$, that is,  $F(x) \in \lbrace 0, 1, \alpha, \cdots, \alpha^{2^{n}-2} \rbrace $ for any $x \in \mathbb{F}_{2^{n}}$, where $\alpha$ is a primitive element of $\mathbb{F}_{2^{n}}$, then 
\begin{center}
$x^{2^{n}-1}+1 =\underset{a \in \mathbb{F}_{2^{n}}}{\prod} gcd (F(x)-a, x^{2^{n}-1}+1)$.
\end{center}
Since that $F^{2^{n}}(x)=F(x) \,mod \, (x^{2^{n}-1}+1)$
\begin{center}
$\dfrac{x^{2^{n}-1}+1}{gcd(F(x), x^{2^{n}-1}+1)} = \underset{a \in \mathbb{F}_{2^{n}}^{*}}{\prod} gcd(F(x)-a, x^{2^{n}-1}+1)$.
\end{center}
Then
\begin{center}
$g(x) \equiv 0 \, mod \,(\underset{a \in \mathbb{F}_{2^{n}}^{*}}{\prod}gcd(F(x)-a, x^{2^{n}-1}+1))$,
\end{center}
hence $g(x)$ is a codeword of the $2^{n}-$ary cyclic code with generator polynomial $G_{F}$.
\end{proof}

The following lemma describes the set of zeros of the cyclic with generator polynomial $G_{F}$ given  by (\ref{vect}) in Theorem \ref{theo1}.

\begin{lemma}\label{c(f)} 
Let $F$ be an $(n,m)$-function in univariate form, the nonzero elements of $F^{-1}(b)$, for $b\in \mathbb{F}_{2^{m}}$, are the zeros of the cyclic code generated by $G_{F}$ denoted by $\mathcal{C}(F^{-1}(b))_{b\in\mathbb F_{2^m}}$.
\end{lemma}
\begin{proof}
Let $g(x) = \overset{2^{n}-1}{\underset{i=0}{\sum}}a_{i}x^{i}$ be an annihilator of $F$, then there exist $b \in \mathbb{F}_{2^{m}}$ such that $g(x)=0$ for every $x \in F^{-1}(b)$. Since $g$ is associated to a codeword of the code generated by $G_{F}$, then for $x \neq 0$, $\overset{2^{n}-1}{\underset{i=1}{\sum}}a_{i}x^{i} = x \overset{2^{n}-1}{\underset{i=1}{\sum}}a_{i}x^{i-1}=0$ that is $\overset{2^{n}-2}{\underset{j=0}{\sum}}a_{j+1}x^{j}=0$. Thus it holds for every codeword of the cyclic code.
\end{proof}

An immediate result from Theorem \ref{theo1} is the following.

\begin{corollary}
The algebraic immunity of an $(n,m)$-function $F$ is equal to the minimal weight-heigh of a codeword of the $2^{n}-$ary cyclic code $\mathcal{C}(F^{-1}(b))_{b\in\mathbb F_{2^m}}$ with generator polynomial $G_{F}$ given by (\ref{vect}).
\end{corollary}
\begin{proof}
Suppose $g \in AN(F)$ with algebraic degree equal to the algebraic immunity of $F$.  Then by Theorem \ref{theo1}, $g$ is associated to a codeword $(a_{1},\cdots,a_{2^{n}-1})$ of $\mathcal{C}(F^{-1}(b))_{b\in\mathbb F_{2^m}}$. Since that the degree of $g$ is equal to $\underset{j=0, \cdots ,2^{n}-1; \, a_{j}\neq 0}{max}wt_{2}(j)$, then for all $a_{j} \neq 0$, such that $0\leq j \leq 2^{n}-1$), $wt_{2}(j)$ is equal to the weight-height of $g$.
\end{proof}

\subsection{Bound on the algebraic immunity of vectorial functions}
In \cite{Mesn}, the authors gave the following lower bound on the $LDA(f)$ by studying the properties of the $2^{n}-$ary cyclic codes $\mathcal{C}(supp(f))$.

\begin{theorem} (\cite{Mesn})\label{Thmesn}
Let $f : \, \mathbb{F}_{2^{n}} \rightarrow \mathbb{F}_{2}$ and $\delta$ be the minimum distance of $\mathcal{C}(supp(f))$, where $e$ is the lowest positive integer such that $\sum_{i=0}^{e} \begin{pmatrix} n \\ i \end{pmatrix} \geq \delta$. Then $LDA(f) \geq e$. \end{theorem}

The next result is an extension of  Theorem \ref{Thmesn} providing a bound on the algebraic immunity of $(n,m)$-functions. 

\begin{theorem}\label{lem1}
Let $F$ be an $(n,m)$-function in univariate form, $b\in \mathbb{F}_{2^{m}}$ and $\delta$ be the minimum distance of $\mathcal{C}(F^{-1}(b))_{b\in\mathbb F_{2^m}}$, hence the following hold.\\ 
$(i)$ If $e$ is the lowest positive integer such that verifying $\underset{i=0}{\overset{e}{\sum}} \begin{pmatrix} n \\ i \end{pmatrix} < \delta$, then $F$ has no nonzero annihilator of algebraic degree less than or equal to $e$.\\
$(ii)$ If $e$ is the lowest positive integer such that verifying $\underset{i=1}{\overset{e}{\sum}}\begin{pmatrix} n \\i \end{pmatrix} \geq \delta$, then $LDA(F) \geq e$.
\end{theorem}
\begin{proof}
Let $g(x)=\underset{i=0}{\overset{2^{n}-1}{\sum}} a_{i}x^{i}$ be an annihilator of $F$ of algebraic degree at most $e$, that is, $a_{i}=0$ for every $i$ of $wt_{2}(i) > e$. 
Let $c$ be a codeword of $\mathcal{C}(F^{-1}(b))_{b\in\mathbb F_{2^m}}$, thus the number of the nonzero components of $c$ is at most $\underset{i=0}{\overset{e}{\sum}} \begin{pmatrix} n \\ i \end{pmatrix}$, that is the weight of $c$ is less than $\delta$. Hence $c$ is the null codeword which prove that any annihilator of $F$ is of algebraic degree at least $e+1$. For the $(ii)$ part, suppose that $e$ is the smallest integer such that $\underset{i=0}{\overset{e}{\sum}}\begin{pmatrix} n \\ i \end{pmatrix} \geq \delta$. Then $\underset{i=0}{\overset{e-1}{\sum}}\begin{pmatrix}n\\i\end{pmatrix} < \delta$ and then $LDA(F) \geq e$. 
\end{proof}

\subsection{Spectral immunity of vectorial functions} 
In the following we give a relation between the spectral immunity and algebraic immunity of an $(n,m)$-vectorial function $F$ and the weight of $\mathcal{C}(F^{-1}(b))_{b\in\mathbb F_{2^m}}$.\\
Let $x \in \mathbb{F}_{2^{n}}$ be the initial state, then the filter generator produce the key-stream $z_{t}$, which is a sequence over $\mathbb{F}_{2}^{n}$, such that $z_{t} = F(x\alpha^{t})$ for $t=0,1,\cdots$. 

Let $g$ be a Boolean function in univariate form, so that $g(x\alpha^{t})=u_{t}$ is an annihilator of $z_{t}$. According to Definition \ref{def1}, the spectral immunity of $z_{t}$ is the lowest linear complexity of $u_{t}$ which is the number of nonzero coefficients in
\begin{center}
$g(x)=\underset{i=0}{\overset{2^{n}-2}{\sum}}\delta_{i}x^{i} \in \mathbb{F}_{2^{n}}[x]$.
\end{center}

\begin{theorem}
Let $F$ be an $(n,m)$-functions function given in its univariate representation.  Then under the hypothesis of Theorem \ref{theo1}, the following hold\\
$(i)$ The spectral immunity of $F$ is equal to the minimal weight of the $2^{n}-$ary cyclic code with generator polynomial $G_{F}$.\\
$(ii)$ The relation between the spectral immunity and the algebraic immunity of $F$ is given by the following inequality
\begin{center}
$SI(F) \leq \underset{i=0}{\overset{AI(F)}{\sum}} \begin{pmatrix} n \\ i \end{pmatrix}.$
\end{center}
\end{theorem}

\begin{proof}
By definition, the algebraic immunity of an $(n,m)$-functions function is the lowest algebraic degree of its nonzero annihilator $g$.
Let $g(x)=\underset{i=0}{\overset{2^{n}-1}{\sum}}a_{i}x^{i}$ be a Boolean function of algebraic degree at least $AI(F)$ associated to a codeword $c=(a_{1}, \cdots, a_{2^{n}-1})$ of $\mathcal{C}(F^{-1}(b))_{b\in\mathbb F_{2^m}}$ such that $a_{i}=0$ for all $i$ with $wt_{2}(i) \leq AI(F)$. Therefore $g$ has at least $\underset{i=0}{\overset{AI(F)}{\sum}} \begin{pmatrix} n \\ i  \end{pmatrix}$ nonzero coefficients in its univariate representation, then $SI(F) \leq \underset{i=0}{\overset{Al(F)}{\sum}}\begin{pmatrix}n \\i \end{pmatrix}$.  
\end{proof}

\section{The algebraic complement of vectorial functions}
In this section, we  present a generalization of the notion of algebraic complement to $(n,m)$-functions. Next we give a bound on the lowest algebraic degree annihilator of $F^{c}$.\\
We start by giving some results about the algebraic complement of Boolean functions.

\subsection{The algebraic complement of Boolean functions}
In \cite{ZhPiZh}, the authors extended the concept of algebraic immunity of Boolean function $f$ and then introduced the notion of algebraic complement $f^{c}$ as given in the following definition
\begin{definition} \cite{ZhPiZh}
Given a Boolean function $f$ on $\mathbb{F}^{n}_{2}$, we define an algebraic complement of $f$, denoted by $f^{c}$, as the function that contains all monomials $x_{1}^{a_{1}} ,\cdots,x_{n}^{a_{n}}$, where each $a_{j} \in \mathbb{F}_{2}$, that are not in ANF of the function $f$. That is any pair of functions $(f, f^{c})$ does not have any monomials in common.
\end{definition}

In \cite{WanChe} and \cite{ZhPiZh} the authors have exhibited  some properties of the algebraic complement of Boolean function.
Set $\Delta (x) = \underset{i=1}{\overset{n}{\prod}}(1+ x_{i})$ where $x=(x_{1},\cdots,x_{n}) \in \mathbb{F}^{n}_{2}$. Then, they have shown that $f(x) \Delta(x) = 0$ for $f(0)=0$ and $f(x) \Delta(x)= \Delta (x)$ for $f(0)=1$. Moreover, they have proved that
 function $f^{c}$ satisfies  $f^{c}=f+\Delta$ (noted that $f^{c}(x)= f(x)$ for all $x \in \mathbb{F}^{n}_{2} \setminus \lbrace0\rbrace$).

The next proposition gives an explicit relationship between $AN(f)$ and $AN(f^{c})$.

\begin{proposition}\cite{WanChe}
Let $f$ be a Boolean function such that $f(0)=1$ then $AN(f^{c})=AN(f) \cup AN(f)^{c}$.
\end{proposition}

\subsection{The algebraic complement of vectorial functions}
The notion of algebraic complement can be extended to $(n,m)$-functions as follows.

\begin{definition}
Let $F=(f_{1}, \cdots, f_{m})$ be an $(n,m)$-vectorial function. The algebraic complement of $F$, denoted by $F^{c}$, is defined as $F^{c}=(f^{c}_{1}, \cdots, f_{m}^{c})$, where $f_{i}^{c}$ is the algebraic complement of $f_{i}$, for $1\leq i\leq m$.
\end{definition}

In the following, we give some properties of $F^{c}$ as well as a relationship between $AN(F)$ and $AN(F^{c})$.
\begin{lemma}
Let $F$ be an $(n,m)$-function, then 
\begin{enumerate}
\item $F^{c}(x)=F(x)$ for all nonzero $x \in \mathbb{F}^{n}_{2}$.
\item $F^{c}(x)=F(x)+\Delta(x)(1, \cdots, 1)$ for all $x \in \mathbb{F}^{n}_{2}$.
\end{enumerate}
\end{lemma}
\begin{proof}
From \cite[Lemma 2]{ZhPiZh} we have $f_{i}(x)=f^{c}_{i}(x)$ for all nonzero $x \in \mathbb{F}^{n}_{2}$, $1 \leq i \leq m$, then for all nonzero $x \in \mathbb{F}^{n}_{2}$ we have that
\begin{equation*}
\begin{split}
F^{c}(x)=(f^{c}_{1}(x), \cdots, f^{c}_{m}(x))
&= (f_{1}(x), \cdots,f_{m}(x)) = F(x).
\end{split}
\end{equation*}
From \cite[Lemma 2]{ZhPiZh} we have for all $x \in \mathbb{F}^{n}_{2}$, $f^{c}(x)= f(x)+\Delta(x)$, then
\begin{equation*}
\begin{split}
F(x)^{c}&=(f^{c}_{1}(x), \cdots, f^{c}_{m}(x))\\
&=(f_{1}(x)+\Delta(x), \cdots, f_{m}(x)+\Delta(x))=F(x)+\Delta(x)(1, \cdots, 1).
\end{split}
\end{equation*}
The proof is completed.
\end{proof}

We recall that a Boolean function $g$ is said to be an annihilator of $F$ if there exists $b \in \mathbb{F}^{m}_{2}$ such that $g(x)=0$ for all $x \in F^{-1}(b) =\lbrace a \in \mathbb{F}^{n}_{2}; \, f_{i}(a)=b_{i},\, 1\leq i\leq m\rbrace$.\\
In the following we give a relation between $F^{-1}(b)$ and $(F^{c})^{-1}(b)$.
\begin{lemma}\label{lem2}
Let $F$ be an $(n,m)$-function and $F^{c}$ its algebraic complement, the following results hold\\
$(i)$ If $0 \in F^{-1}(b)$ then $(F^{c})^{-1}(b)=F^{-1}(b) \cap \mathbb{F}^{n}_{2}\setminus \lbrace 0 \rbrace$.\\
$(ii)$ If $0 \notin F^{-1}(b)$ then $(F^{c})^{-1}(b)=F^{-1}(b) \cup \lbrace 0 \rbrace$.\\
$(iii)$ If $0 \notin F^{-1}(b)$ and $0 \notin (F^{c})^{-1}(b)$ then $F^{-1}(b) = (F^{c})^{-1}(b)$.
\end{lemma}

\begin{proof}
Let $g \in AN(F)$, $g(x)=0$ for all $x \in F^{-1}(b)$. From \cite[Lemma 2, item 1]{ZhPiZh} we have $f_{i}(x)=f^{c}_{i}(x)$ for a nonzero $x \in \mathbb{F}^{n}_{2}$, thus $(F^{c})^{-1}(b) = F^{-1}(b)$ for all nonzero $a \in \mathbb{F}^{n}_{2}$.\\
Let $0 \in F^{-1}(b)$, then $0 \notin (F^{c})^{-1}(b)$, therefore $(F^{c})^{-1}(b)=F^{-1}(b) \cap \mathbb{F}^{n}_{2} \setminus\lbrace 0 \rbrace$.\\
Let $0 \notin F^{-1}(b)$, then $0 \in (F^{c})^{-1}(b)$, therefore $(F^{c})^{-1}(b)=F^{-1}(b) \cup \lbrace 0 \rbrace$.
\end{proof}

\begin{lemma}\label{lem3}
Let $g \in AN(F)$, with the previous notations set $AN(F)^{c}= \lbrace g^{c}, \, g \in AN(F)\rbrace$, then\\
$(i)$ $g \in AN(F^{c})$ for $F^{-1}(b)=(F^{c})^{-1}(b)$.\\ 
$(ii)$ $AN(F) \subset AN(F^{c})$ with $0 \in F^{-1}(b)$.\\
$(iii)$ $AN(F)^{c} \subset AN(F^{c})$ with $0 \notin F^{-1}(b)$.\\

Let $g \in AN(F^{c})$, then\\ 
$(iv)$ $AN(F^{c}) \subset AN(F)$ for $0 \notin F^{-1}(b)$.\\
$(v)$ $AN(F^{c})^{c} \subset AN(F)$ for $0 \in F^{-1}(b)$.
\end{lemma}
\begin{proof}
Let $g$ be an annihilator of $F$, for a given $b \in \mathbb{F}^{m}_{2}$, $g(x)=0$ for all $x \in F^{-1}(b)$.\\
First suppose $0 \notin F^{-1}(b)$ and $0 \notin (F^{c})^{-1}(b)$, then by item $(iv)$ of Lemma \ref{lem2}, $F^{-1}(b)=(F^{c})^{-1}(b)$, which gives $g(x)=0$ for all $x \in F^{-1}(b)=(F^{c})^{-1}(b)$, then $g$ is an annihilator of $F^{c}$. Now if $0 \in F^{-1}(b)$ then $0 \notin (F^{c})^{-1}(b)$, from item $(iii)$ of Lemma \ref{lem2}, $F^{-1}(b)=(F^{c})^{-1}(b) \cup \lbrace0\rbrace$. Then $g(x)=0$ for all $x \in (F^{c})^{-1}(b) \cup \lbrace0\rbrace$. Thus $g(x)=0$ for all $x \in (F^{c})^{-1}(b)$. So $g \in AN(F^{c})$ which gives $AN(F) \subset AN(F^{c})$.\\
Since that $0 \notin F^{-1}(b)$, then from item $(iii)$ of Lemma \ref{lem2}, $F^{-1}(b)=(F^{c})^{-1}(b) \cap \mathbb{F}^{n}_{2}-\lbrace0\rbrace$, which gives $g(x)=0$ for all $x \in (F^{c})^{-1}(b) \cap \mathbb{F}^{n}_{2}\setminus\lbrace0\rbrace$. So $g^{c} \in AN(F^{c})$ since that $g^{c}$ takes the same value of $g$ for all nonzero $x \in \mathbb{F}^{n}_{2}$ and the value $0$ for $x=0$. So $AN(F)^{c} \subset AN(F^{c})$.
\end{proof}

The relation between the annihilators of the $(n,m)$-function $F$ and the annihilators of its Algebraic complement is given by the following theorem.

\begin{theorem}
Let $F$ be an $(n,m)$-function and $F^{c}$ its algebraic complement, then\\
$(i)$ $AN(F^{c}) = AN(F) \cup AN(F)^{c}$ for $0 \in F^{-1}(b)$.\\
$(ii)$ $AN(F) = AN(F^{c}) \cup AN(F^{c})^{c}$ for $0 \notin F^{-1}(b)$.\\
$(iii)$ $AN(F) = AN(F^{c})$ for $F^{-1}(b)=(F^{c})^{-1}(b)$.
\end{theorem}

\begin{proof}
Let $A_{F}=\lbrace g \in AN(F),$ $0 \in F^{-1}(b)\rbrace \subset AN(F)$, $A_{F^{c}}=\lbrace g \in AN(F^{c}),$ $0 \in (F^{c})^{-1}(b) \rbrace \subset AN(F^{c})$ and let $0 \in F^{-1}(b)$, $g \in AN(F)$. This means that $g(x)=0$ for all $x \in F^{-1}(b)$, then $g(0)=0$, therefore $A_{F} = AN(F)$.\\
By item $(i)$ of Lemma \ref{lem3}, $g \in A_{F^{c}}$ for all $g \in A_{F}$ which gives $A_{F^{c}} \subset A_{F}$. By item $(iii)$ of Lemma \ref{lem3}, $g \in A_{F}$ for all $g \in A_{F^{c}}$, then $A_{F} \subset A_{F^{c}}$, this implies that $A_{F}=AN(F)=A_{F^{c}}$.\\
Let $B_{F^{c}}=\lbrace g \in AN(F^{c})$, $0 \notin (F^{c})^{-1}(b) \rbrace$, then $\#AN(F^{c})= \#A_{F^{c}} + \#B_{F^{c}}$. By item $(ii)$ of Lemma \ref{lem3}, $g \in A_{F}$ then $g^{c} \in B_{F^{c}}$. But $A_{F}=A_{F^{c}}$, then $g \in B_{F^{c}}$ for all $g \in A_{F^{c}}$. Thus $\#A_{F^{c}} = \#B_{F^{c}}$. \\
Let $\varphi$ be a mapping from $A_{F^{c}}$ to $B_{F^{c}}$, such that for all $p \in A_{F^{c}}$, $\varphi(p)=p^{c} \in B_{F^{c}}$. To prove that $\varphi$ is injective we take $p_{1}^{c},$ $p_{2}^{c}$ such that $p_{1}^{c}=p_{2}^{c}$, then $p_{1}+\Delta(x)=p_{2}+\Delta(x)$ this implies that $p_{1}=p_{2}$.\\
Since $\#A_{F^{c}}=\#B_{F^{c}}$, then $\varphi$ is surjective. Then we have that $AN(F^{c})=A_{F^{c}} \cup B_{F^{c}}= AN(F) \cup AN(F)^{c}$. Which proves the theorem's first item. When $0 \in F^{-1}(b)$, $0 \notin (F^{c})^{-1}(b)$. We prove item $(ii)$ of the Theorem in same way as item $(i)$.\\
Let $F^{-1}(b)=(F^{c})^{-1}(b)$. By item $(v)$ of Lemma \ref{lem3}, if $g \in AN(F)$ then $g \in AN(F^{c})$, thus $AN(F^{c}) \subset AN(F)$, if $g \in AN(F^{c})$ then $g \in AN(F)$, thus $AN(F) \subset AN(F^{c})$. That gives $AN(F)=AN(F^{c})$, which completes the proof. 
\end{proof}

\subsection{On the lowest algebraic degree of nonzero annihilators of  Boolean and vectorial functions }

In this subsection, we are firstly interested on the lowest algebraic degree of nonzero annihilators of  Boolean functions.

Let $\alpha$ be a primitive element of $\mathbb{F}_{2^{n}}$, we define the set of $t$ consecutive zeros of $\mathcal{C}(supp(f))$ by $V(\alpha, l, t)= \lbrace \alpha^{l}, \alpha^{l+1}, \ldots, \alpha^{l+t-1} \rbrace$.

\begin{theorem}\label{theo2}\cite{MacSlo}
Let $\alpha$ be a primitive element of $\mathbb{F}_{2^{n}}$ and let $r$, $k$ and $t$ be non-negative integers, $m$ a positive integer relatively prime to $n$.
Let $\mathcal{C} \subset \mathbb{F}_{2^{n}}$ a cyclic code having $\alpha^{r}, \alpha^{r+1}, \ldots, \alpha^{r+t-1}, \alpha^{r+m}, \alpha^{r+m+1}, \ldots, \alpha^{r+m+t-1}, \ldots, \\ \alpha^{r+km}, \alpha^{r+km+1}, \ldots, \alpha^{r+km+t-1}$.
Then the minimum distance of $\mathcal{C}$ $\delta$ is greater then $t+k$.
\end{theorem}

In \cite{Mesn1}, the following result was proved.
\begin{theorem}\label{20}
Let $f$ : $\mathbb{F}_{2^{n}} \rightarrow \mathbb{F}_{2}$ be a Boolean function and $l$, $k$ and $\delta$ be positive integers, let $m$ be a positive integer which is relatively prime to $n$. Suppose $S_{f}$ contains $V(\alpha,l,\delta-1) \cup V(\alpha,l+m,\delta-1) \cup \ldots \cup V(\alpha,l+km,\delta-1)$. Then $LDA(f) \geq e$, where $e$ is the smallest positive integer such that $\underset{i=1}{\overset{e}{\sum}}\begin{pmatrix}n \\ i \end{pmatrix}\geq \delta+k$.     
\end{theorem}

In following we give a lower bound for $LDA(1+f)$, where $f$ is defined as in Theorem \ref{20}.

\begin{theorem}\label{mm}
Let $f$ : $\mathbb{F}_{2^{n}} \rightarrow \mathbb{F}_{2}$ be a Boolean function, let $l$, $k$ and $\delta$ be positive integers, $m$ be a positive integer relatively prime to $n$. Then\\ $(i)$ $Supp(1+f)$ contains $V(\alpha, l+\delta -1, 2^{n}-\delta)\cup V(\alpha, l+m+ \delta -1, 2^{n}- \delta) \cup \ldots \cup V(\alpha, l+km+\delta -1, 2^{n} -\delta)$.\\
$(ii)$ $LDA(1+f)$ $\geq e-1$, where $e$ is the smallest possible positive integer such that $\underset{i=1}{\overset{e}{\sum}} \begin{pmatrix} n \\ i \end{pmatrix} \geq 2^{n}-\delta +k$.
\end{theorem}

\begin{proof}
Let $f$ be a Boolean function and $\mathcal{C}(supp(f))$ the cyclic code with generator polynomial $G_{f}$.\\
$supp(1+f) = zeros(f) \cap \mathbb{F}_{2^{n}}^{*}$, then
\begin{equation*}
\begin{split}
    S_{1+f} & = \lbrace 1, \alpha, \ldots, \alpha^{l-1}, \alpha^{l+\delta-1},\ldots, \alpha^{2^{n}-2} \rbrace \cup\lbrace 1, \alpha, \ldots, \alpha^{l+m-1}, \alpha^{l+m+\delta-1},\ldots, \alpha^{2^{n}-2} \rbrace  \\
    & \quad \cup \ldots\cup \lbrace 1, \alpha, \ldots, \alpha^{l+km-1}, \alpha^{l+km+\delta-1},\ldots, \alpha^{2^{n}-2} \rbrace.\\
    & = \lbrace \alpha^{l+\delta-1}, \ldots, \alpha^{l+\delta-1+(2^{n}-\delta)-1} \rbrace \cup \lbrace \alpha^{l+m+\delta-1},\ldots,\alpha^{l+m+\delta-1+(2^{n}-\delta)-1} \rbrace \cup \ldots \\
    & \quad \cup \lbrace \alpha^{l+km+\delta-1}, \ldots, \alpha^{l+km+\delta-1+(2^{n}-\delta)-1} \rbrace.\\
    & =V(\alpha, l+\delta-1, 2^{n}-\delta)\cup V(\alpha, l+m+\delta-1, 2^{n}-\delta) \cup \ldots \cup V(\alpha, l+km+\\
    & \quad \delta-1, 2^{n}-\delta).
\end{split}
\end{equation*}
Thanks to \cite[Corollary 02]{Mesn} and Theorem \ref{theo2}, $LDA(1+f) \geq e-1$, where $e$ is the smallest positive integer such that $\underset{i=1}{\overset{e}{\sum}} \begin{pmatrix} n \\ i \end{pmatrix} \geq 2^{n}-\delta +k$.
\end{proof}

As a consequence, one can immediately derive a bound on the algebraic immunity of Boolean function.

\begin{corollary}
Let $f$ and $1+f$ be defined as in Theorem \ref{20} and Theorem \ref{mm}, respectively. Then $AI(f) \geq e-1$, where $e$ is the smallest integer such that $\underset{i=1}{\overset{e}{\sum}}\begin{pmatrix} n \\ i \end{pmatrix} \geq min(\delta +k, 2^{n}-\delta +k)$.
\end{corollary}

Now, we are interested in vectorial functions. The following result gives a link between the lowest algebraic degree of nonzero annihilators of vectorial functions and of its algebraic complement.

\begin{proposition}\label{born}
Let $F$ be an $(n,m)$-function and $F^{c}$ be its algebraic complement. Then we have
\begin{center}
$LDA(F)-1 \leq LDA(F^{c}) \leq LDA(F)+1$.
\end{center}
\end{proposition}

\begin{proof}
Let $g$ be an annihilator of $F$ of degree $LDA(F)$. Since that $F(x)= F^{c}(x)$ for all $x \in \mathbb{F}^{*}_{2^{n}}$, $xg(x)F^{c}(x)=0$ for all $x \in \mathbb{F}_{2^{n}}$. Then $LDA(F^{c}) \leq LDA(F)+1$.\\
Let $g$ be an annihilator of $F^{c}$, clearly $xg(x)F(x)=0$ for all $x \in \mathbb{F}_{2^{n}}$. Then $LDA(F)-1 \leq LDA(F^{c})$.
\end{proof}

\section{Further properties of the cyclic $\mathcal{C}(F^{-1}(b))_{b\in\mathbb F_{2^m}}$ associate  to the vectorial function $F$ }
In this section we firstly prove that the constructed cyclic code $\mathcal{C}(F^{-1}(b))_{b\in\mathbb F_{2^m}}$ is a linear code with complementary dual (abbreviated LCD). Recall that LCD codes are linear codes whose intersection with their dual are trivial \cite{mas92}.
When they are binary, they play an important role in armoring implementations against side-channel attacks and fault injection attacks \cite{CG14}.
In following, we recall some basic results on LCD cyclic codes \cite{Mas}. 

\begin{definition}(Self reciprocal polynomial)
The polynomial $G(x)$ will be called self-reciprocal if and only if $G(x)=G_{0}^{-1}x^{deg(G)}G(x^{-1})$.
\end{definition}

\begin{proposition}\cite{Mas}\label{prop}
Let $\mathcal{C}$ be a cyclic code generated by a monic polynomial $G(x)$, then the following statements are equivalent.\\
$(i)$ $\mathcal{C}$ is an LCD code.\\
$(ii)$ $\mathcal{C}$ is generated by the self-reciprocal polynomial $G(x)$.
\end{proposition}

As shown in Theorem \ref{theo1}, for certain $b \in \mathbb{F}_{2}^{m}$, the annihilators of an $(n,m)$-function in univariate form belong to the $2^{n}-$ary cyclic code $\mathcal{C}(F^{-1}(b))_{b\in\mathbb F_{2^m}}$ with generator polynomial $G_{F}$. Let $\mathcal{C}^{\perp}$ be the dual of $\mathcal{C}(F^{-1}(b))_{b\in\mathbb F_{2^m}}$. Then we can deduce easily the generator polynomial of $\mathcal{C}^{\perp}$.

\begin{corollary}
The dual code $\mathcal{C}^{\perp}$ of the cyclic code $\mathcal{C}(F^{-1}(b))_{b\in\mathbb F_{2^m}}$ is generated by the polynomial
\begin{center}
$G^{\perp}_{F} = x^{deg(G)}G(x^{-1})$,
\end{center}
where $G(x)= gcd(F(x), x^{2^{n}-1}+1)$.
\end{corollary}

In other words, the generator polynomial of $\mathcal{C}^{\perp}$ is the reciprocal polynomial of $G(x)$.\\
By definition, the zeros of $\mathcal{C}^{\perp}$ are the inverses of the non zeros of $\mathcal{C}(F^{-1}(b))_{b\in\mathbb F_{2^m}}$, so from Lemma \ref{c(f)}, are the inverses of the elements of zero of $F$.

In the following, we present the parameters of the LCD $\mathcal{C}(F^{-1}(b))_{b\in\mathbb F_{2^m}}$ code.

\begin{theorem} 
For a given $b \in \mathbb{F}_{2^{m}}$, $\mathcal{C}(F^{-1}(b))_{b\in\mathbb F_{2^m}}$ is an LCD cyclic code with parameters $[2^{n}-1,\,2^{n-1}-\dfrac{1}{2}W_{\phi_{b}}(0),\,d]$.
\end{theorem}

\begin{proof}
Let $\beta_{i}$ ($\beta_{i}$ be a primitive element of $\mathbb{F}_{2^{n}}$) be a zero of $G(x)$, then $\beta^{-1}_{i}$ is a zero of $G_{F}^{\perp}(x)$. Which implies that $G(\beta_{i})= G_{F}^{\perp}(\beta^{-1}_{i})=0$. Thus $G(x)=G_{F}^{\perp}(x)$ for all $x \in \lbrace x\in \mathbb{F}_{2^{n}},\,F(x)=0 \rbrace$, this implies that $\lbrace \beta_{1}, \ldots, \beta_{r} \rbrace = \lbrace \beta^{-1}_{1}, \ldots, \beta^{-1}_{r} \rbrace$. Then from Proposition \ref{prop}, $\mathcal{C}(F^{-1}(b))_{b\in\mathbb F_{2^m}}$ is an LCD code.\\
By definition, the degree of the generator polynomial of $\mathcal{C}(F^{-1}(b))_{b\in\mathbb F_{2^m}}$ is equal to $|D|$, where $D$ the defining set of $\mathcal{C}(F^{-1}(b))_{b\in\mathbb F_{2^m}}$. From Lemma \ref{c(f)}, $D$ is the set $F^{-1}(b)$ for $b$ in $\mathbb{F}_{2^{m}}$.\\
Let $\phi_{b}$  be the indicator function of $F^{-1}(b)$ defined over $\mathbb{F}_{2^{n}}$ by $\phi_{b}(x)=1$ if $F(x)=b$ and $\phi_{b}(x)=0$ otherwise. Then $|F^{-1}(b)|= |\lbrace x \in \mathbb{F}_{2^{n}},\, \phi_{b}(x)=1 \rbrace|$. Hence,    
\begin{equation*}
    \begin{split}
        |D|&=|\lbrace x\in \mathbb{F}_{2^{n}},\,\phi_{b}(x)=1\rbrace|\\
        &=2^{n}-|\lbrace x\in \mathbb{F}_{2^{n}},\,\phi_{b}(x)=0\rbrace|\\
        &=2^{n}-\dfrac{1}{2}\underset{y\in\mathbb{F}_{2}}{\sum}\,\,\underset{x\in \mathbb{F}_{2^{n}}}{\sum}(-1)^{y\phi_{b}(x)}\\
        &=2^{n-1}-\dfrac{1}{2}W_{\phi_{b}}(0).
    \end{split}
\end{equation*}
Then the dimension of $\mathcal{C}(F^{-1}(b))_{b\in\mathbb F_{2^m}}$ is equal to $2^{n-1}-\dfrac{1}{2} W_{\phi_{b}}(0)$.
\end{proof}

Since the set of annihilators of an $(n,m)$-function containing the annihilators of its algebraic complement (the inverse is true), then we give the link between the cyclic codes associated to $F$ and $F^{c}$ as follows.

\subsubsection{Link between $\mathcal{C}(F^{-1}(b))_{b\in\mathbb F_{2^m}}$ and $\mathcal{C}((F^{c})^{-1}(b^{'}))_{b^\prime\in\mathbb F_{2^m}}$}
Let $F$ be an $(n,m)$-function in univariate form and $F^{c}$ be its algebraic complement. In the following we give a relation between $\mathcal{C}(F^{-1}(b))_{b\in\mathbb F_{2^m}}$ and $\mathcal{C}((F^{c})^{-1}(b^{'}))$, for some  $b,\,b^{'} \in \mathbb{F}_{2^{m}}$.

\begin{lemma}
Let $F \,: \, \mathbb{F}_{2^{n}} \rightarrow \mathbb{F}_{2^{m}}$ and $F^{c}$ its algebraic complement, $b,\,b^{'} \in \mathbb{F}_{2^{m}}$. Let $\mathcal{C}(F^{-1}(b))_{b\in\mathbb F_{2^m}}$ and $\mathcal{C}((F^{c})^{-1}(b^{'}))$ be the $2^{n}-$ary cyclic codes generated by $G_{F}$ and $G_{F^{c}}$, then \\
$(i)$ $\mathcal{C}(F^{-1}(b)) \subset \mathcal{C}((F^{c})^{-1}(b^{'}))$.\\
$(ii)$ $\exists \, h \in \mathbb{F}_{2^{n}}[x]$ such that $G_{F}=G_{F^{c}}.h$.
\end{lemma}

\begin{proof}
Let $0 \in F^{-1}(b)$, which means that $0 \notin (F^{c})^{-1}(b^{'})$. Since $F^{c}(x)=F(x)$ for all $x \in \mathbb{F}_{2^{n}}^{*}$ then $F^{-1}(b)=(F^{c})^{-1}(b^{'})\cap \mathbb{F}_{2^{n}}^{*}$.\\
Suppose that $g \in AN(f)$ associated to a codeword $(a_{1}, \ldots, a_{2^{n}-1})$ of $\mathcal{C}(F^{-1}(b))_{b\in\mathbb F_{2^m}}$, then $\underset{i=0}{\overset{2^{n}-1}{\sum}}a_{i}x^{i}=0$ for all $x \in F^{-1}(b)$, this implies that $a_{0}=0$ for all $x \in F^{-1}(b)$. Thus $\underset{i=1}{\overset{2^{n}-1}{\sum}}a_{i}x^{i}=0$ for all $x \in F^{-1}(b) \cap \mathbb{F}^{*}_{2^{n}}$. Then $(a_{1}, \ldots, a_{2^{n}-1}) \in \mathcal{C}((F^{c})^{-1}(b^{'}))$. Thanks to \cite[Theorem 2.29, item 1]{Cun}, $G_{F^{c}}$ divide $G_{F}$, so it exists a polynomial $h(x) \in \mathbb{F}_{2^{n}}[x]$ such that $G_{F} = G_{F^{c}}.h(x)$.
\end{proof}

\subsubsection{The weight distribution of $\mathcal{C}(F^{-1}(b))_{b\in\mathbb F_{2^m}}$}
The non zeros weights of the LCD code associated to an $(n,m)$-function $F$ are given as below. First, we recall that an upper bound on the algebraic immunity of $(n,m)$-functions was given in \cite{ArmKra} in the next lemma.
\begin{lemma} \label{upper}
For an $(n,m)$-function $F$, $Al(F) \leq d$ where $d$ is the smallest integer such that $\underset{i=0}{\overset{d}{\sum}}\begin{pmatrix} n\\i \end{pmatrix} > 2^{n-m}$. 
\end{lemma}

By considering the case where $n=m$, the following proposition holds.
\begin{proposition}\label{upper1}
Let $F$ be an $(n,m)$-function, then $Al(F) \leq 1$. 
\end{proposition}

\begin{proof}
From Lemma \ref{upper}, $Al(F)\leq d$ where $d$ is the smallest integer such that $\underset{i=0}{\overset{d}{\sum}}\begin{pmatrix} n\\i \end{pmatrix} > 2^{n-m}$. Let $n=m$, then $d$ is the smallest integer such that $\underset{i=0}{\overset{d}{\sum}}\begin{pmatrix} n\\i \end{pmatrix} > 2^{n-n}=1$, that is $d=1$ and $Al(F)\leq 1$. 
\end{proof}

In the following, we give the non-zeros weights of the LCD code .
\begin{theorem}
Let $F$ be an $(n,m)$-function and $\mathcal{C}(F^{-1}(b))_{b\in\mathbb F_{2^m}}$ of minimal distance $\delta$. Let $e$ be the smallest integer such that $\underset{i=0}{\overset{e}{\sum}}\begin{pmatrix} n\\i \end{pmatrix}\geq \delta$, then 
\begin{center}
$e\leq Al(F)\leq 1$ 
\end{center}
and we have \\
$(i)$ If $e=1$, then $\mathcal{C}(F^{-1}(b))_{b\in\mathbb F_{2^m}}$ has codewords of non zero weights $\geq 1+n$.\\ 
$(ii)$ If $e=0$, then $\mathcal{C}(F^{-1}(b))_{b\in\mathbb F_{2^m}}$ is the trivial LCD code.
\end{theorem}

\begin{proof}
Suppose that $g=\underset{i=0}{\overset{2^{n}-1}{\sum}}$ is an annihilator of $F$ of algebraic degree $LDA(F)$, $g$ is associated to a codeword $c$ of $\mathcal{C}(F^{-1}(b))_{b\in\mathbb F_{2^m}}$, then $wt(c)$ is equal to $\underset{i=0}{\overset{LDA(F)}{\sum}}\begin{pmatrix} n\\i \end{pmatrix}$.\\
From Theorem \ref{lem1}, $LDA(F) \geq e$ where $e$ is the smallest integer such that $\underset{i=0}{\overset{e}{\sum}}\begin{pmatrix} n\\i \end{pmatrix}\geq \delta$ and from Proposition \ref{upper1} we have that $e \leq Al(F) \leq 1$. Thus $e \in \lbrace 0,1\rbrace$.\\
If $e=1$, then $wt(c) \geq \underset{i=0}{\overset{1}{\sum}}\begin{pmatrix} n\\i \end{pmatrix}=1+n$.\\ 
If $e=0$, the result is obvious.
\end{proof}

\section{Conclusions}
In this paper a direct link between the annihilators of $(n,m)$-functions (in univariate form) and certain $2^{n}-$ary cyclic codes was given. The knowledge of the minimum distance of those codes give rise to a lower bound on the algebraic immunity of the $(n,m)$-functions.
Some properties of cyclic codes (whose generator polynomials were determined by vectorial functions) as well as their weight enumerator were given. Further, we gave a generalization and some properties of the so-called algebraic complement as well as a lower bound on the lowest algebraic degree of nonzero annihilators.


\begin{thebibliography}{99}

\bibitem{ArmKra}F. Armknecht and M. Krause.: "Constructing single-and multi-output Boolean functions with maximal immunity", ICALP (2), pp. 180-191, 2006.


\bibitem{Bat}L.M. Batten.: Algebraic attacks over GF(q), Lecture Notes in Computer Science 3348, Springer, ISBN 3-540-24130-2, pp. 84-91, 2004.



\bibitem{CarlB} C. Carlet, "Boolean Functions for Cryptography and error-correcting codes", Boolean Models and Methods in Mathematics, Computer Science, and Engineering. Y. Crama, P.L. Hammer. pp. 257-397, 2010.

\bibitem{Carl}C. Carlet.: Vectorial Boolean Functions for Cryptography, Boolean Models and Methods in Mathematics, Computer Science, and Engineering. Y. Crama, P.L. Hammer. pp. 398-470, 2010.

\bibitem{CaDaGuMa}C. Carlet, D.K. Dalai, K.C. Gupta and S. Maitra.: Algebraic immunity for cryptographically significant Boolean functions : Analysis and construction, IEEE Trans. Inform. Theory 52(7), pp. pp. 3105-3121, 2006.

\bibitem{CarlFre} C. Carlet and K. Feng.: An infinite class of balanced functions with optimal algebraic immunity, good immunity to fast algebraic attacks and good nonlinearity, In ASIACRYPT, LNCS 5350. Springer-Verlag, pp. 425-440, 2008.


\bibitem{CG14} C. Carlet and S. Guilley.: Complementary dual codes for counter-measures to side-channel attacks, In: E. R. Pinto et al. (eds.), Coding Theory and Applications, CIM Series in Mathematical Sciences, vol. 3, pp. 97-105, Springer Verlag, 2014 and Journal Adv. in Math. of Comm. 10(1), pp. 131-150, 2016.

\bibitem{Nico} N. Courtois.: Algebraic attacks on stream ciphers, Extended slides by Courtois. www.nicolascourtois/papers/toyolili-slides.pdf.

\bibitem{CourMei}N. Courtois and W. Meier.: Algebraic attacks on stream ciphers with linear feedback, In Advances cryptology-Eurocrypt 2003. Lecture Notes in computer science. Springer-Verlag, vol. 2656, pp. 345-359, 2003.

\bibitem{CourM}N. Courtois and W. Meier.: Algebraic attacks on stream ciphers with linear feedback, extended version of the Eurocrypt 2003.

\bibitem{CourPie}N. Courtois and J. Pieprzyk.: Cryptanalysis of block ciphers with over-defined systems of equations, Proceedings of ASIACRYPT 2002, Lecture Notes in Computer Science 2501, pp. 267-287, 2003.

\bibitem{Cun} C. Ding.: Codes from Difference sets, World Scientific Publishing Company, 2014.

\bibitem{Feng} K. Feng, Q. Liao and J. Yang.: Maximal values of generalized algebraic immunity, Designs, Codes and Cryptography 50(2), pp. 243-252, 2009. 

\bibitem{GoHellRon}G. Gong, S. R\o njom, T. Helleseth and H. Hu.: Fast Discrete Fourier Spectra Attacks on stream Ciphers, IEEE Trans. Inform. Theory 57(8), pp. 5555-5565, 2011.


\bibitem{HellRon}T. Helleseth and S. R\o njom.: Simplifying Algebraic Attacks with Univariate Analysis, In Proceedings of Information Theory and Applications Workshop (ITA), pp. 1-7, 2011.


\bibitem{Liu-Mesnager} J. Liu, S. Mesnager, L. Chen.:
On the nonlinearity of S-boxes and linear codes. Journal Cryptography and Communications 9(3): 345-361, 2017.

\bibitem{MacSlo} F.J. MacWilliams and N.J. Sloane.: The theory of error correcting codes, Amsterdam, North Holland 1977.
\bibitem{Mas}J.L. Massey, "Reversible codes", Information\&Control, Vol. 7. pp. 369-380, Sept 1964.


\bibitem{mas92}  J. L. Massey.: Linear codes with complementary duals, Discrete Math., vol. 106-107, pp. 337-342, 1992.


\bibitem{Mats} M. Matsui.:  Linear cryptanalysis method for DES cipher, advences in cryptology-EUROCRYPT 1993. LNCS, vol. 765, Springer-verlag, 1993, pp. 386-397. 

\bibitem{MeiPasCar} W. Meier, E. Pasalic and C. Carlet.: Algebraic attacks and decomposition of Boolean functions, in Eurocrypt 2004. Lecture notes in computer science, volume 3027, 2004.

\bibitem{Mesn} S. Mesnager, G.D. Cohen.: Cyclic codes and algebraic immunity of Boolean functions, ITW: pp. 1-5, 2015.

\bibitem{Mesn1} S. Mesnager.: A note on linear codes and algebraic immunity of Boolean functions, 21-st International Symposium on Mathematical Theory of Networks and Systems July 7-11, 2014.

\bibitem{NN} K. Nyberg.: Perfect nonlinear S-Boxes, in advences in cryptology-EUROCRYPT, New York, NY, USA, Springer- Verlag, pp. 378-385, 1991.


\bibitem{SHAN} C.E. Shannon.: Communication theory of secrecy systems, Bell system technical journal, 28, pp. 656-715, 1949.
.

\bibitem{WanChe} C.P. Wang and X.S. Chen.: On extended algebraic immunity", Des. Codes Cryptography 57(3), pp. 271-281, 2010.

\bibitem{WaChZh} J. Wang, K. Chen and S. Zhu.: Annihilators of Fast Discrete Fourier Spectra Attacks," Advances in Information and Computer security, 7-th international workshop on security, IWSEC 2012 Fokuoka, Japan, pp. 182-196, 2012.

\bibitem{ZhPiZh} Xian-Mo Zhang, J. Pieprzyk and Y. Zheng.: On algebraic Immunity and Annihilators, Lecture note in computer Science, Information Security and Cryptology-ICISC, pp. 65-77, 2006.

\bibitem{ZhSoDuWe} J. Zhang, S. Song, J. Du and Q. Wen.: On the construction of multi-output Boolean functions with optimal algebraic immunity, science China Information Sciences 55(7), pp. 1617-1623, 2012.

\end{thebibliography}
\end{document}